\documentclass[aps,twocolumn,prl,color,psfig,epsf,sort&compress,floatfix]{revtex4-2}
\usepackage{amssymb}
\usepackage{amsmath}
\usepackage{bm}
\usepackage[dvipsnames]{xcolor}
\usepackage{soul}
\usepackage{hyperref}
\usepackage{graphicx}
\usepackage{etoolbox}
\usepackage{tikz}
\newrobustcmd*{\mysquare}[1]{\tikz{\filldraw[draw=#1,fill=#1] (0,0)
rectangle (0.2cm,0.2cm);}}
\begin{document}

%\title{Thin blue fog and finite quasi-crystal blue phase patterning in non-Euclidean topologies}
\title{Cholesteric shells: two-dimensional blue fog and finite quasicrystals}

\author{L.N. Carenza$^{1,2}$, G. Gonnella$^1$, D. Marenduzzo$^3$,  G. Negro$^{1*}$, E.  Orlandini$^4$}
\affiliation{
$^1$Dipartimento di Fisica, Universit\'a degli Studi di Bari and INFN, Sezione di Bari, via Amendola 173, Bari, I-70126, Italy, \\
$^2$ Instituut-Lorentz, Universiteit Leiden, P.O. Box 9506, 2300 RA Leiden, Netherlands,\\
$^3$ SUPA, School of Physics and Astronomy, University of Edinburgh, Peter Guthrie Tait Road, Edinburgh, EH9 3FD, UK \\
$^4$ Dipartimento di Fisica e Astronomia, Università di Padova, 35131 Padova, Italy \\
$^*$ Corresponding Author: giuseppe.negro@ba.infn.it
}

\begin{abstract}
We study the phase behaviour of a quasi-two dimensional cholesteric liquid crystal shell. We characterise the topological phases arising close to the isotropic-cholesteric transition, and show that they differ in a fundamental way from those observed on a flat geometry. For spherical shells, we discover two types of quasi-two dimensional topological phases: finite quasicrystals and amorphous structures, both made up by mixtures of polygonal tessellations of half-skyrmions. These structures generically emerge instead of regular double twist lattices because of geometric frustration, which disallows a regular hexagonal tiling of curved space. For toroidal shells, the variations in the local curvature of the surface stabilises heterogeneous phases where cholesteric patterns coexist with hexagonal lattices of half-skyrmions. Quasicrystals, amorphous and heterogeneous structures could be sought experimentally by self assembling cholesteric shells on the surface of emulsion droplets.
\end{abstract}

\maketitle

%\section{Introduction}
Chiral liquid crystals have remained of high interest to physicists for decades, because they simultaneously provide a fertile ground for applications to nanotechnology~\cite{bluephaselasers,skyrmionssmalyukh,hopfionsmalyukh,Schwartz2018} as well as a practical realisation of topological phases in condensed matter~\cite{Wright89,selinger,knotscolloidslc}. A paradigmatic example is that of blue phases, which arise close to the isotropic-cholesteric transition and consist of 3D packings of double twist cylinders~\cite{Wright89}. These phases have been used as lasers~\cite{bluephaselasers} or display devices~\cite{kikuchibpdevice}, and proposed as templates for colloidal photonic crystals~\cite{mihacolloids}. Blue phases I and II are crystalline, %\textcolor{black}{-like}, 
whereas the structure of blue phase III, also called the blue fog, has long constituted a puzzle in condensed matter physics~\cite{Wright89}. Early theories predicted it to be either a quasicrystal or an amorphous solid. More recent computer simulations~\cite{henrich2011} and photopolymerisation experiments~\cite{kentbp3} have showed that the latter model is more accurate, and suggest that the blue fog is a thermodynamically stable amorphous lattice of disclinations, locally akin to blue phase II. 

Chiral liquid crystals also form hexagonal lattices of double twist cylinders, called half-skyrmions or merons, in thin quasi-2D samples, and arrays of ring defects or more exotic knotted field states, known as hopfions, in thicker samples~\cite{Fukuda2011,Nych2017,Metselaar2019,Fukuda2011_b,smalyukh}. Hopfions and half-skyrmions are topological quasi-particles which can be created optically and manipulated by an electric field~\cite{smalyukh}. Blue phases and half-skyrmion lattices arise due to a phenomenon known as {\it topological frustration}: the chiral nature of the underlying molecules locally favours doubly twisted structures, but double twist cylinders create director field patterns which cannot be patched together smoothly without creating defects, or disclination lines. The structures seen in experiments and predicted theoretically are therefore those which provide the best compromise between the favourable double twist and the energetically costly defects~\cite{Rokhsar86,Nych2017}. 
%\textcolor{black}{Moreover, recent experiments on shells of cholesteric liquid crystals have shown that a plethora of different surface structures can be obtained as the result of the competition of chirality and soft confinement, including cholesteric stripes and a patching of focal conic domains\textcolor{black}{~\cite{Tran2017}}.}

Here we use lattice Boltzmann simulations to study what phases form when a cholesteric liquid crystal is confined to a~thin shell surrounding a {\it curved} closed surface, whose width is less than a cholesteric pitch. Henceforth we refer to this system as a cholesteric shell.
%. Our study complements those in~\cite{Tran2017,Darmon2016,Durey2020}, which instead mainly focussed on much thicker shells, with width larger or much larger than the pitch, where cholesteric stripes or focal domain patterns appear~\cite{Tran2017}.
Their experimental realization was studied in~\cite{Tran2017,Darmon2016,Durey2020} which  instead mainly focused on much thicker shells, with width larger or much larger than the pitch, where cholesteric stripes or focal domain patterns appear~\cite{Tran2017}.

%We mostly focus on the case of spherical shells, which can be more readily investigated experimentally~\cite{rassegnadifettiliquidcrystalsgocce}. 
For spherical shells, we discover that the topological phases emerging close to the isotropic-cholesteric transition are fundamentally different from the regular hexagonal lattices of half-skyrmions found for flat geometries~\cite{selinger,Metselaar2019,Duzgun2020,Duzgun2021}. 
The curved geometry of spherical shells introduces an additional geometric frustration, as the Gauss-Bonnet theorem forces the total topological defect charge of the tessellation to equal the Euler characteristic of the surface which is $+2$ for spherical topologies~\cite{ireth2021,rassegnadifettiliquidcrystalsgocce,Pollard2019}. As a result, a regular hexagonal lattice of half-skyrmions is impossible to realise, as its overall topological charge is $0$.
At small radii, we observe the formation of finite quasicrystals which consist of polygonal mixtures reminiscent of the structures formed by patchy colloids on the surface of a droplet~\cite{glotzerquasicrystals}. As the radius of the confining shell is increased, these regular structures give space to amorphous arrangements with a multifarious variety of double twist polygons scattered with no rule: they may be viewed as an analogue of the blue fog in a curved 2D geometry. Close to this topological transition we observe {\it{scars}} --chains of alternated pentagons and heptagons-- which mediate the loss in regularity of the tessellation.

%or amorphous ones where a multifarious variety of double twist polygons tessellate the surface. Quasicrystals are favoured for smaller shell radii and  Amorphous structures are more complicated and show no underlying tessellation rule: they may be viewed as an analogue of the blue fog in a curved 2D geometry. %, and, like it, they appear to be thermodynamically stable in a range of parameters.

We also find that curvature can direct pattern formation and self-assembly in shells with non-constant curvature (like a torus). This feature can be exploited  to tilt the balance in favour of either helical patterns or a regular half-skyrmion lattice, %respectively favoured by negative and positive curvature, and 
resulting in heterogeneous systems. % This results in heterogeneous systems, or geometrically driven phase coexistence.
%\textcolor{blue}{We also find that curvature can be further exploited to direct pattern formation and self-assembly in cholesteric shells. For instance, the non-constant curvature of a torus can be exploited  to tilt the balance in favour of a helical pattern (cholesteric phase, favoured in regions of negative curvature), or of a regular half-skyrmion lattice (favoured in regions of positive curvature). This results in heterogeneous systems, or geometrically driven phase coexistence.} %For sufficiently large chirality, the tendency to twist dominates and the shell arranges into a homogeneous half-skyrmion lattice, as in the plane. 
This rich phase behaviour could be probed experimentally with cholesteric shells of variable curvature. 
%\Livio{The groundbreaking nature of our study is twofold as we bridge two striking topics in Soft Matter --\emph{i.e.} the development of exotic topological phases and ordering in curved geometries-- by connecting the effects of topological frustration in terms of the well-known Thomson problem and vice versa providing a physical realization of topology-induced morphological transition in liquid-crystalline systems.}
Our study bridges the topics of topological frustrations in cholesterics with that of ordering on a closed geometry. The case of spherical shells can also be viewed as a generalisation of the Thomson problem -- finding the optimal arrangements of point-like charges on a sphere -- to phase-shifting topological quasiparticles (half-skyrmions which can attain the form of any polygon).

We use a Landau-de Gennes approach to model a cholesteric shell, constituted by a chiral liquid crystal (LC), with orientational order described by the nematic tensor $\bm{Q}$. \textcolor{black}{To stabilize the LC shell we confine $\bm{Q}$ to a thin interface of a fluid droplet, described by a phase field $\phi$ for computational convenience.} The free-energy of the system is $\mathcal{F}=\mathcal{F}^{chol} + \mathcal{F}^{\phi}$, where
\begin{subequations}\label{eq:hydrodynamics} 
\begin{gather}
\begin{split}
    \mathcal{F}^{chol}=  \int \text{d}\bm{r} \textcolor{black}{\bigg\lbrace} A_0 \left[ \dfrac{1}{2}  \left(1 - \dfrac{\chi}{3} \right)\bm{Q}^2 -  \dfrac{\chi}{3} \bm{Q}^3 +  \dfrac{\chi}{4} \bm{Q}^4 \right]  \\ + \dfrac{L}{2}\left[\left( \nabla \cdot \bm{Q} \right)^2 + \left(\nabla \times \mathbf{Q} + 2 q_0 \bm{Q} \right]^2 \right] \textcolor{black}{\bigg\rbrace},
\end{split} \label{eqn:freeEchol}\\ 
\mathcal{F}^{\phi}=  \int \text{d}\bm{r} \left[ \frac{a}{4} \phi^2 (\phi-\phi_0)^2 + \frac{k_\phi}{2} (\nabla \phi)^2 \right].
\end{gather}
\end{subequations}

%\st{Our model for a cholesteric shell is constituted by a chiral liquid crystal (LC), with orientational order described by a symmetric and traceless tensor $\bm{Q}$, confined to a thin interface defined in terms of a phase field $\phi$. The free energy $\mathcal{F}$  of the system consists of three parts. The first, describing  cholesteric ordering, is}
%environment with total density $\rho=1$ and divergence-free flow field $\bm{v}$. The properties of the equilibrium state are defined by the following free energy density 
%\begin{multline}
%\mathcal{F}^{chol}=  \int \text{d}\bm{r} A_0 \left[ \dfrac{1}{2}  \left(1 - \dfrac{\chi}{3} \right)\bm{Q}^2 -  \dfrac{\chi}{3} \bm{Q}^3 +  \dfrac{\chi}{4} \bm{Q}^4 \right]  \\ + \dfrac{L}{2}\left[\left( \nabla \cdot \bm{Q} \right)^2 + \left(\nabla \times \mathbf{Q} + 2 q_0 \bm{Q} \right]^2 \right].
%
%\end{multline}
The term proportional to the energy scale $A_0$ describes the isotropic-cholesteric transition which occurs at $\chi>\chi_{cr}=2.7$, 
whilst the one proportional to the elastic constant $L$ accounts for the energy cost of elastic deformations. % in the single constant approximation.
The parameter $q_0>0$ favors right-handed twist with equilibrium pitch $p_0=2\pi q_0^{-1}$ in bulk systems~\cite{Carenza22065,carenza2020_physA}.
%The second part of the free energy 
%$\mathcal{F}^{\phi}=  \int \text{d}\bm{r} \left[ \frac{a}{4} \phi^2 (\phi-\phi_0)^2 + \frac{k_\phi}{2} (\nabla \phi)^2 \right]
%\label{eqn:freeEphi}
%$
%settles the shell geometry. %phase field structures. 
For $a>0$ there are two possible equilibrium values for $\phi$ ($0$ and $\phi_0$) while $k_\phi$ determines the surface tension and the interface width. % $\xi = (4 k_\phi/a \phi_0)^{1/2}$. 
To model a spherical shell, we create a droplet of radius $R$ in the phase field ($\phi\simeq \phi_0$ inside and $\phi\simeq 0$ outside) and confine the LC to the interface by setting $\chi=\chi_0 + \chi_s (\nabla \phi)^2$, with $\chi_0 < \chi_{cr}$, so that the $Q$-tensor is different from $0$ only on a thin shell (width $\sim \xi$) at the droplet interface (Fig.~S1)~\cite{Metselaar2019b,Liu2003,Lee2012}. 
Toroidal shells can be obtained through a suitable spacial patterning of $\phi_0(\bm{r})$~\cite{SI}.
\textcolor{black}{}
%\textcolor{blue}{To simulate spherical shells, we create a droplet (radius $R$) in the phase field, so that $\phi\simeq \phi_0$ inside the droplet and $\phi\simeq 0$ outside, while the interface is associated with large values of $(\nabla \phi)^2$. To create a cholesteric shell, we confine the chiral LC to the interface by setting $\chi=\chi=\chi_0 + \chi_s (\nabla \phi)^2$ in Eq.~\eqref{eqn:freeEchol} (in our simulations $\chi_0= 10 \chi_s = 2.5$).} 
 %Finally, tangential anchoring is imposed through the additional free energy $\mathcal{F}^{anch}= W \int \text{d}\bm{r}   Q_{\alpha \beta} \partial_\alpha \phi \partial_\beta \phi,
%$
%with $W>0$~\textcolor{black}{\cite{note_anchoring}}.
\begin{figure}[t]
	\centerline{\includegraphics[width=0.98\columnwidth]{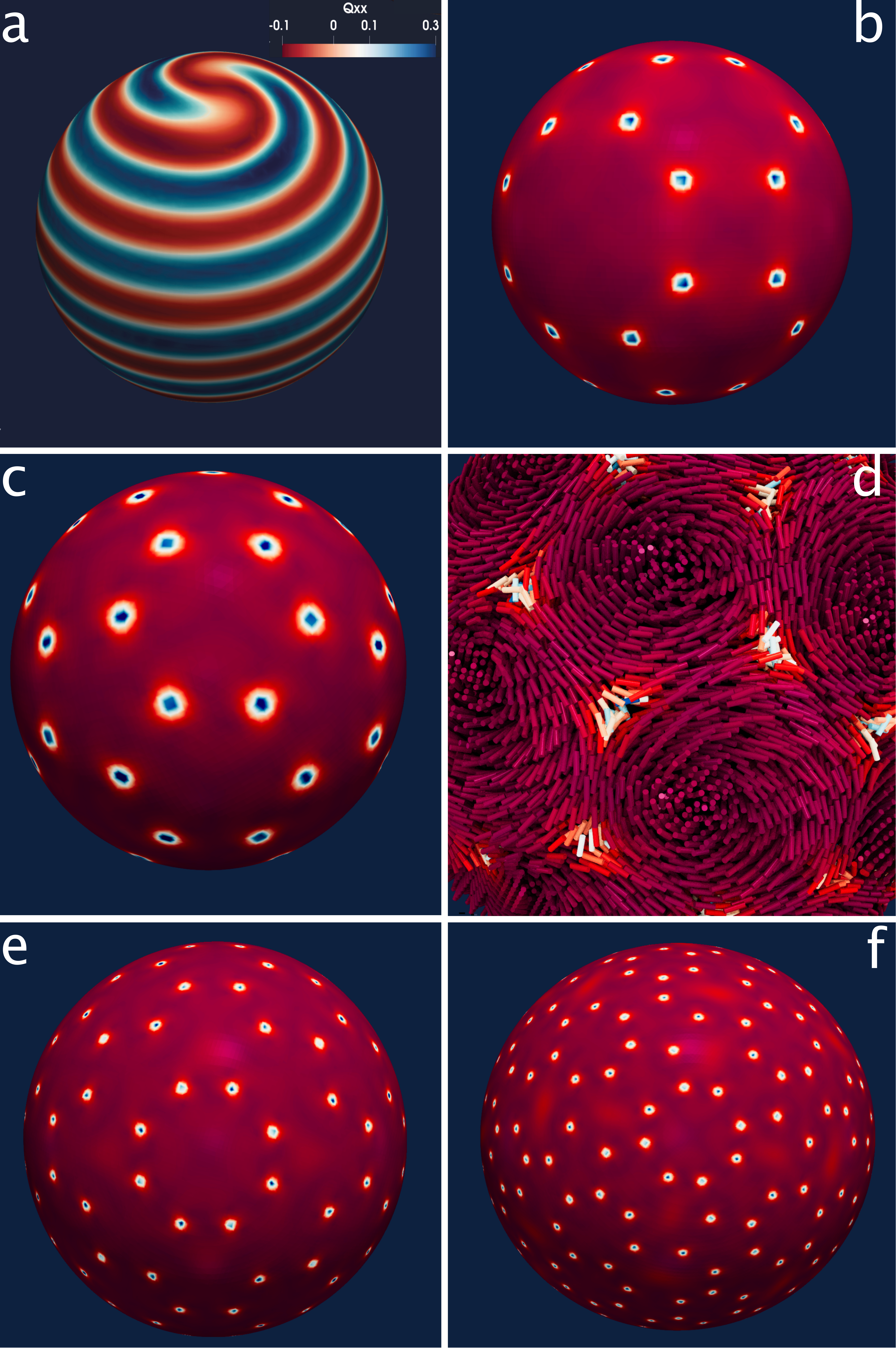}}
	\caption{\textbf{Shell configurations.}  (a) Contour-plot of $Q_{xx}$ in the helical phase at $R=40,L=10^{-3}$.  Three \textit{fqc} configurations are shown in panels~(b-c-e), respectively \textit{OHS} at $R=40,L=10^{-2}$, \textit{HP} at $R=40, L=3 \times 10^{-2}$ and \textit{OHP} at $R=50,L= 2 \times 10^{-2}$ (see text for acronyms).  Panel~(d) shows the director field pattern of two half-skyrmions and nine $-1/2$ defects defining a pentagon and a hexagon in panel~(c).
	Panel~(f) shows an amorphous configuration at $R=80, L=4 \times 10^{-2}$.
	Color code in panel~(b-f) corresponds to the isotropy parameter $c_s$ of the Westin metrix~\cite{callan2006,SI}: blue regions define defect positions ($\bm{Q} \sim 0$), while red ones are ordered ($\bm{Q} \neq 0$).
%	The frame color of each panel has been chosen to match the corresponding color code used for the phase diagram in Fig.~\ref{fig3}(f).
	}\label{fig1}
\end{figure}

The key control parameters of the system are: \emph{(i)} the reduced temperature $\tau={9(3-\chi)}/{\chi}$, \emph{(ii)} the chirality strength $\kappa=\sqrt{108 q_0^2 L /(A_0 \chi)}$ proportional to the ratio between nematic coherence length and cholesteric pitch~\cite{alexander2009,Wright89}, and \emph{(iii)} the ratio between shell radius and cholesteric pitch, $R/p_0$~\cite{depablo2015}. We set parameters such that the interfacial thickness $\xi = (4 k_\phi/a \phi_0)^{1/2}\ll p_0$, to model thin shells. % \Livio{so that the twisting is always directed in the tangent plane of the shell rather than radially.}
%\begin{equation}
%  \tau = \dfrac{9(3-\chi)}{\chi} \quad \text{and} \quad \kappa = \sqrt{\dfrac{108 q_0^2 L }{A_0 \chi}}.
%  \label{eqn:temp_chir}
%\end{equation} 
In the bulk, Eq.~\eqref{eqn:freeEchol} is minimised by the helical phase~\cite{Wright89}, for $\tau<\tau_c(\kappa)=\frac{1}{8} \left[1-4 \kappa^2 + \left(1+ 4 \kappa^2/3\right)^{3/2}  \right]$. 
%\textcolor{blue}{In the bulk, Eq.~(\ref{eqn:freeEchol}) is minimised by the helical phase (N*), characterized by unidirectional twist with equilibrium pitch $p_0=2\pi q_0^{-1}$, for $\tau<\tau_c(\kappa)=\frac{1}{8} \left[1-4 \kappa^2 + \left(1+ 4 \kappa^2/3\right)^{3/2}  \right]$.} 
The isotropic phase is stable if $\tau\gtrsim 0.8 \tau_c ({\kappa=0})$. In 3D, blue phases (BP) are found for sufficiently large chirality between the helical and isotropic phase~\cite{braz1975_b}. In 2D
%, \textcolor{blue}{\sout{in the same parameter region}} 
regular half-skyrmion lattices with hexagonal symmetry appear~\cite{Metselaar2019}.

LC hydrodynamics is ruled by a set of time-dependent differential equations~\cite{SI}. The Beris-Edwards equation $D_t \bm{Q} = \bm{H}$, where $D_t$ is the material derivative for a tensor field and the molecular field $\bm{H}=-\frac{\delta \mathcal{F}}{\delta \bm{Q}}+ \frac{\bm{I}}{3} Tr \left(\frac{\delta \mathcal{F}}{\delta \bm{Q}} \right)$ %(with $\mathcal{F}=\mathcal{F}^{chol}+ \mathcal{F}^{\phi}+\mathcal{F}^{anch}$) 
drives the LC towards its equilibrium state~\cite{SI}. 
The phase field $\phi$ evolves according to a Cahn-Hilliard-like equation ~\cite{SI}. \textcolor{black}{We stress that the model parameters are chosen in such a way that the droplet does not deform, so that the volume occupied by the LC is factually conserved during the relaxation dynamics.}
Finally, the Navier-Stokes equation for the flow field $\bm{v}$ accounts for momentum balance with an elastic stress depending on the orientational order. 
The inclusion of hydrodynamic interactions lowers the likelihood for the system to get trapped into metastable states~\cite{henrich2011}. The equations are solved via a hybrid lattice Boltzmann approach~\cite{denniston2001,orlandini2008,succi2001,carenza2019,bonelli2019} in 3D grids of size ranging from $128^3$ to $384^3$ with periodic boundary conditions. 
Further simulation details and parameters are given in~\cite{SI}.

%\section{Quasi-crystalline and amorphous textures in shell-confined BP}
We start from the case of spherical shells. We fix $\tau=0.540$, $q_0=0.245$, and vary $\kappa$ and $R/p_0$ (which we controlled by modifying $L$ and $R$ respectively). 
%The chirality $\kappa$ was controlled in droplets with different radii by varying the elastic constant $L$ of the liquid crystal in accordance to Eq.~\eqref{eqn:temp_chir}  which, with our choice of parameters reads $\kappa \simeq 6.61 L^{1/2}$.
For low $\kappa$ %elastic constants $L \lesssim 0.008$ chirality
($\kappa \lesssim 0.12, L \lesssim 10^{-3}$) the system is in the helical phase where the LC arranges into a spiral pattern winding around the shell. The spiral axis is defined by two pairs of $+1/2$ defects at each pole of the shell (Fig.~\ref{fig1}(a)). For high chirality ($\kappa \gtrsim 1.1, L \gtrsim 0.07$), the system is in the isotropic phase. %The defect pattern is reminiscent of that observed with colloidal particle with tangential anchoring embedded in a cholesteric liquid crystal.

\begin{figure}[t]
	\centerline{\includegraphics[width=1.0\columnwidth]{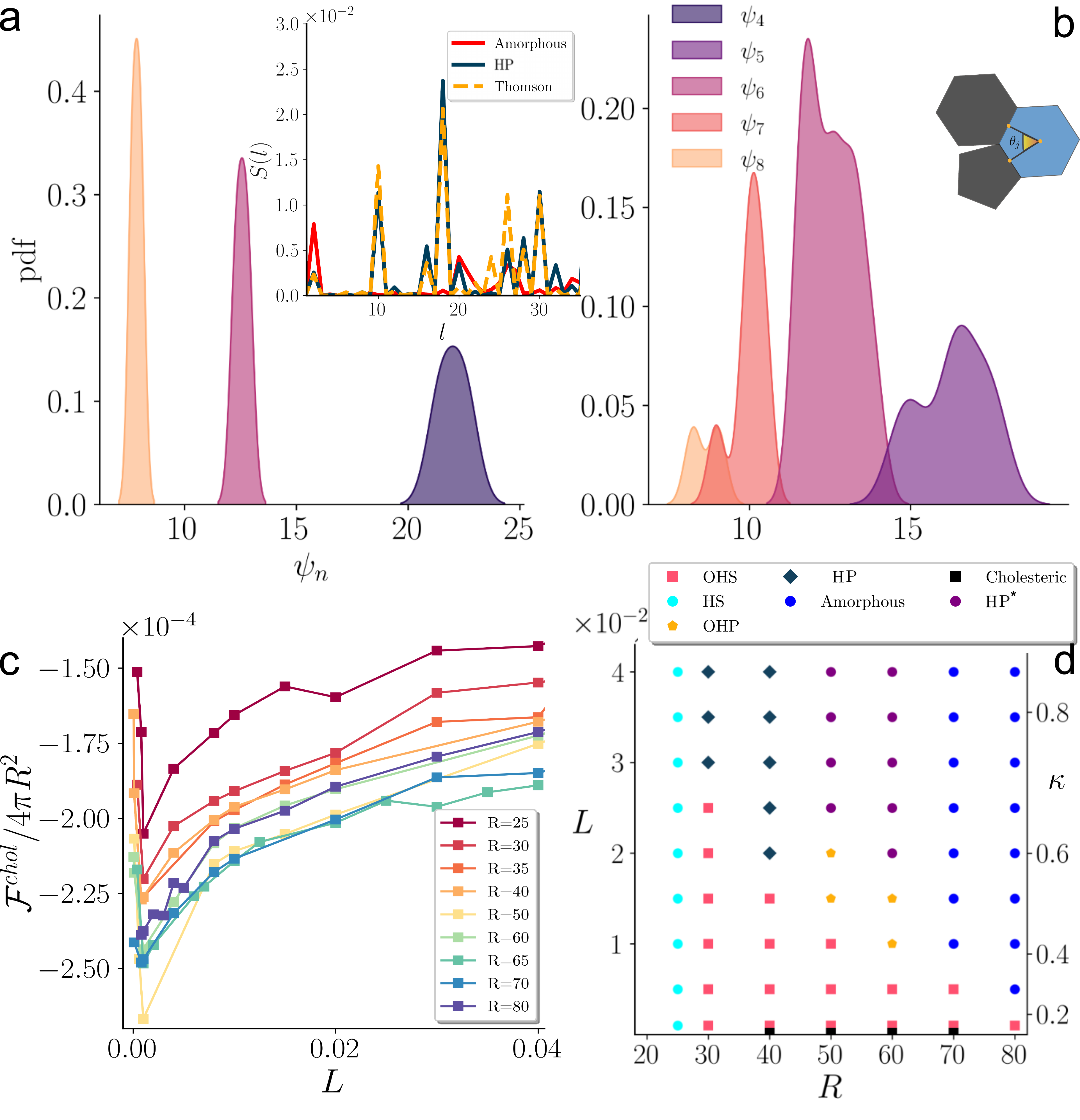}}
	\caption{\textbf{Finite quasi-crystals and amorphous configurations.} 
	%(a) \textcolor{blue}{Schematic representation of the polyhedron net a throuple of neighboring polygons in the plane} considered to compute the order parameter $\psi_n$. (b-c) $\psi_n$ distribution for a fqc and an amorphous configuration respectively shown in Fig.~\ref{fig1}(b-f). Notice that the expected value for the OHS fqc in panel~(b) are $\psi_4=22, \psi_6=12.6, \psi_8=7.85$ (see SI). (d) Distribution of the distance between first-neighbor defects for the fqc and amorphous configuration. (e) Free energy vs. elastic constant $L$ for some radii. (f) Phase diagram of observed configurations in the $L-R$ plane. In the legend, letters in the acronyms stand for $S$=square, $P$=penthagons, $H$=hexagons, $O$=octagons.}\
	(a-b) Normalized $\psi_n$ \emph{pdf} for the \emph{OHS} and the amorphous configuration of Fig.~\ref{fig1}(b) and (f). The expected values are $\psi_4=22, \psi_6=12.6, \psi_8=7.85$~\cite{SI}. The inset shows Bessel spectra for the $HP$ configuration (blue), the analytical solution of the Thomson problem with $32$ particles (dashed yellow), and an amorphous configuration (red).	Inset of panel (b) illustrates the definition of %the order parameter
	$\psi^j_n$. (c) Free energy density vs. $L$. (d) Phase diagram in the $L-R$ plane. In the legend, %letters in the acronyms mean:
	\emph{S}=square, \emph{P}=pentagons, \emph{H}=hexagons, \emph{O}=octagons.}
	\label{fig3}
\end{figure}

For intermediate chirality, topological phases arise. %, as in 2D and 3D. 
In our curved geometry, these emerge as  polygonal tessellations of the surface corresponding to half-skyrmions, separated by point defects with topological charge $-1/2$.  %\textcolor{blue}{ Importantly, hexagonal half-skyrmion lattices as those observed in a flat 2D geometry~\cite{selinger,metselaar2018} have zero total topological charge (as each half-skyrmion region counts as a non-singular, or escaped, defect with topological charge $+1$~\cite{garethcholestericdefects}), and are therefore forbidden by the constraint posed by the Gauss-Bonnet theorem  which states that the total defect charge of a generic liquid crystal pattern on a curved manifold must be equal to the Euler characteristic of the latter, which is $2$ for a sphere. Thus, the ensuing tesselation needs to involve polygons other than hexagons to satisfy this constraint. }
Importantly, hexagonal half-skyrmion lattices as those observed in a flat 2D geometry~\cite{selinger,Metselaar2018} have zero total topological charge, and are therefore forbidden on a sphere by the Gauss-Bonnet theorem~\cite{Kamien2002,rassegnadifettiliquidcrystalsgocce}. Thus, the ensuing tessellation needs to involve polygons other than hexagons. % for topological reasons. 
As an $n$-edges polygon in the tessellation contributes a charge of $1-n/6$,
%\textcolor{black}{($+1$ from the half-skyrmion inside the polygon, and $-1/2 \times 1/3$ for each vertex)}. 
the topological constraint provided by the Gauss-Bonnet theorem can be expressed through the Euler formula for polyhedra as a condition on the number and types of polygons used for the tessellation, $\sum \left( 1- n/6\right)N_n=2$,
where $N_n$ is the number of $n$-edges polygons. %\footnote{Note that this relation is equivalent to the celebrated Euler formula for polyhedra, $V-E+F=2$, as $V=(\sum_n N_n n)/3$, $E=(\sum_n N_n n)/2$ and $F=\sum_n N_n$.}

Fig.~\ref{fig1}(b-f) shows a gallery of different topological phases %, or equivalently half-skyrmion polyhedric configurations,
found at varying both $\kappa$ and $R/p_0$; Fig.~S6 in~\cite{SM} show corresponding predictions for cross-polarized textures which could be observed in experiments. We observe two main types of structures. For small radii (up to $R \simeq 65$) the structures are locally regular, although the tessellations involve a mixture of different polygons. 
%\Livio{Sia nel testo che nelle caption facciamo riferimento ad R=35 che però non mostriamo più nel diagramma di fase... dobbiamo mettere R=30 o 40} 
At $R=40$ and low chirality ($\kappa = 0.424, L=0.01$) we observe a regular network of octagons, hexagons, and squares (which we denote with {\it{OHS}}) where each polygon borders an equal number of polygons of different types in a well-defined orderly fashion: for instance, in an octagon, if an edge borders a hexagon, the neighbouring edges need to border squares (Fig.~\ref{fig1}b). \textcolor{black}{At larger chirality $\kappa = 0.735$ ($L=0.03$), we find a football (or soccer ball) configuration (panel~(c)) composed by hexagons and $12$ pentagons (\emph{HP}), with each hexagon bordering exactly $3$ pentagons --namely a truncated icosahedron}.

%whereas at $R=60$ a regular mixture of octagons, hexagons and pentagons (\emph{OHP}) is observed (see Fig.~\ref{fig1}(e)). %For larger radii (e.g., for $R=60$ in Fig.~\ref{fig4_12}, a tessellation consisting of octagons, hexagons and pentagons ($OHP$) may appear. 
The local order found in this regime is reminiscent of that of quasicrystals~\cite{Scacchi202,Subramanian2016}, hence we refer to these polygonal mixtures as {\it finite quasicrystals} (\emph{fqc}). 
%In this case, the resulting state can be obtained through the repetition of a \emph{unit cell}. For instance, at $R=35$ we observe the formation of a regular network of octagons, hexagons and penthagons (which we denote with the acronym of $OHS$) where each polygons borders with  an equal number of polygons of the different kind in an alternated fashion (for instance, in an octagon, if an edge borders with a hexagon, the neighboring edges border with squares). At $R=38$ the observed state develops a football configuration composed by hexagons and $12$ penthagons ($HP$), with each hexagon bordering $3$ penthagons. For larger radii (see for instance $R=60$ in Fig.~\ref{fig4_12}) polygonal tassellation consisting of octagons, hexagons and penthagons ($OHP$) may appear.
For larger radii, the topological phases are fundamentally different. Polygonal tessellations found in steady state appear much more disordered, and no simple correlation between the types of neighbouring polygons is seen: the resulting half-skyrmion arrangement is instead akin to an amorphous lattice. %, and may be viewed as the 2D analogue of blue phase III, hence we name this phase the \emph{two-dimensional blue fog}. 

To quantify the %extent of positional ordering of defects in the structures, or equivalently the 
degree of regularity of a polygonal tessellation, we introduce a phenomenological order parameter $\psi_n$, where $n$ refers to the component of $n$-edge polygons in the tessellation, defined as follows.
% \emph{fqc} configurations and in the amorphous state we introduced a suitable order parameter $\psi_n$, capable of capturing the degree of regularity of a certain polyhedron. 
Let us denote by $\theta^j$ the angle defined by the midpoints of a pair of neighboring edges and the centre of the corresponding $n$-edged polygon, with $j$ a label identifying the pair. Additionally, let us call $\mathcal{N}^{j}_{1}$ and $\mathcal{N}^{j}_{2}$ the number of edges of the two grey bordering polygons in the inset of Fig.~\ref{fig3}(b). The order parameter $\psi_n^j$ is defined for each pair of neighboring edges $j$ as
\begin{equation}
\psi_n^j = \theta^j (\mathcal{N}^{j}_{1} + \mathcal{N}^{j}_{2}) .
%\quad \text{and} \quad \psi_n=\sum_{n',j} \psi_{n'}^{j} \delta_{nn'}
\label{eqn:psi_n}
\end{equation}
%where $n$ denotes the number of edges of the polygon to which the pair belongs.
%This quantity measures the regularity of the tessellation. \textcolor{green}{ENZO:Questo l'abbiamo già detto . Forse questa frase si può togliere} %since it takes into account both the regularity of each polygon and their neighboring configuration.
%
%This quantity provides a measure of the regularity of both each polygon and the tessellation as a whole. 
%
%For a regular lattice, $\psi^j_n$ does not depend on either $j$ or $n$ (in this case $\mathcal{N}^{j}_{1,2}=n$ and $\theta^j=2\pi/n$) so that its distribution is a Dirac delta functions peaked at $4\pi$~\cite{SI}. For a \emph{fqc}, $\psi_n^j$ still does not depend on $j$, but $\mathcal{N}^j_{1,2}$ may differ from $n$ (as different types of polygons contribute to the tessellation), so the expected distribution is now a Dirac comb with appropriate weights. % instead of a single Dirac delta function. 
For a regular lattice, its distribution $\psi_n$ is a Dirac delta function peaked at $4\pi$ while for a \emph{fqc} is a Dirac comb~\cite{order_parameter,SI}. 
%For a \emph{fqc}, $\psi_n^j$ still does not depend on $j$, but $\mathcal{N}^j_{1,2}$ may differ from $n$ (as different types of polygons contribute to the tessellation), so the expected distribution. % instead of a single Dirac delta function. 
In contrast, for amorphous structures $\psi_n$ should broaden and the peaks flatten.
%In contrast, for amorphous lattices the distribution of $\psi^j_n$ should broaden and the peaks flatten.

Fig.~\ref{fig3}(a) shows $\psi_n$ for the \emph{OHS} configuration of Fig.~\ref{fig1}(b). The distributions are strongly peaked,  %on the predicted values (see caption), 
thus signalling the local regularity expected for a \emph{fqc} state. [The observed spreading is due to a slight deformation of the polygons.]
Instead, the distributions computed for configurations in the large $R$ regime (panel~(b)) are qualitatively different, and spread out over a much wider range of values, thereby we call these structures amorphous~\footnote{For a rigorous identification we should characterise the hexatic order of these tessellations, which requires very large systems to achieve a good accuracy~\cite{digregorio2021}.}.  
The presence of a fundamental difference between these phases is confirmed by an analysis of the Bessel spectra~\cite{copar2019,shtools} of the polygonal tessellations (Fig.~\ref{fig3}a, inset, and~\cite{SI}). The spectrum of the \emph{HP} configuration matches that of the solution of the Thomson problem for the optimal location of charged particles on a sphere, whereas the amorphous state spectrum is less regular.
%This is reinforced by the qualitative difference in the typical spherical structure factor in the two cases (see SI). %structure factor (see SI)
%Therefore, both the distributions of our order parameter $\psi_n$ and Bessel spectral analysis signal a fundamental difference between the \emph{fqc} and amorphous regimes. 

The panoply of possible configurations in the \emph{fqc} regime can be related to the multiple candidate structures arising when minimising the free energy.
%\textcolor{blue}{Indeed, the total free energy of the system can be approximated as (see SI for details)
%$
%f_{\lbrace N_n \rbrace} = \sum_n \left[ f_n  + n f_d/3 \right] N_n ,
%$
%where $f_d$ is the energy ($\sim L$) of a $-1/2$ disclination and $f_n(l)$ %represents the energy associated to a polygon with $n$ edges of length $l$ %which can be computed as
%$
%f_n = \int_{\mathcal{P}_n} \mathit{dS} \mathcal{F}[Q],
%$
%with $\mathcal{P}_n$ the area of an $n$-edge regular polygon, 
%and $\mathcal{F}[Q] \sim r^{-1}$.}
The latter can be approximated as~\cite{SI,Duzgun2018}
$
\mathcal{F}_{\lbrace N_n \rbrace} = \sum_n \left[ \mathcal{F}_n(l)  + n \mathcal{F}_d/3 \right] N_n ,
$
where $\mathcal{F}_d$ is the energy ($\sim L$) of a $-1/2$ disclination and $\mathcal{F}_n(l)=\int_{\mathcal{P}_n(l)}  f[Q] \mathit{dS}$ represents the free energy associated to a polygon $\mathcal{P}_n(l)$ with $n$ edges of length $l$ (with the free energy density $f[Q] \sim 1 + r^{-2}$).
%with $\mathcal{F}[Q] \sim r^{-1}$.
Since the area of each polygon is $A(\mathcal{P}_n)=n \cot\left(\frac{\pi}{n} \right) \frac{l^2}{4}~n^2$, it is energetically favourable to have a large number of polygons with many edges.
However, the more edges, the more negative is the topological charge associated to the polygon, which requires more polygons with a number of edges $n<6$ to satisfy the Euler formula. It is this competition between energy and topology which gives rise to a large variety of possible quasicrystals. %Therefore the equilibrium configuration is expected to exhibit polygons with a number of edges close to 6, but not \emph{too far} from it.

Reasoning along similar lines, one also expects that as $R$ increases, locally different tessellations can be patched together to yield an amorphous structure at only moderate cost, as the density of structural defects arising in the patching should decreases with size. 
Close to the transition between the football configuration and the amorphous phase, we also observe intermediate structures where pentagonal disclinations nucleate lines of dislocations (joint pairs of pentagons and heptagons with null topological charge). These are denoted by \emph{HP*} in Fig.~\ref{fig3}(d), and are similar to scars found in spherical colloidal crystals~\cite{Bowick2000,Bausch2003,Irvine2010}. Scar formation may therefore mediate the transition to our amorphous state.
For sufficiently large $R$, amorphous states are either metastable~\cite{SI} or thermodynamically stable, in which case their free energy is lower than that of any of the quasicrystal phases observed in our simulations (see Fig.~\ref{fig3}(c-d)  and Section~6 of~\cite{SI}). %(see Fig.~\ref{fig3}c  phase diagram in~\ref{fig3}d).
%[Of course, it might be that there are more complicated regular quasicrystals which we have not yet found and which thermodynamically outcompete the 2D blue fog thermodynamically.]). 
For $R\sim 70-80$, we also find an $OHS$-amorphous transition which is triggered by increasing  the elastic constant $L$ (or the LC chirality). These observations suggest that the amorphous phase we have found has properties similar to that of blue phase III~\cite{henrich2011}, or the blue fog, and for that reason we call it the \textit{two-dimensional blue fog}.
%\textcolor{black}{This process can be seen as the analogue in a quasi-crystalline context of \emph{scar screening}~\cite{Bowick2000,Bausch2003,Irvine2010} in two-dimensional spherical crystals, where large strains introduced by the topologically required defects can be alleviated by the proliferation of excess dislocations with null topological charge when the radius $R$ of the spherical surface significantly exceeds the lattice spacing~\cite{GarciaAguilar2020}.}

%\textcolor{black}{The phase diagram of observed configurations in the $L-R$ plane is shown in Fig.~\ref{fig3}d. Metastability was also checked through a series of quenching tests for several values of $L$ and $R$, starting from various initial conditions~\cite{SI}.Importantly, we found that starting from an amorphous configuration and setting the elastic constant below the $OHS$-amorhpous transition at $R \geq 70$, a metastable amorphous region is found even if the free-energy minimum is achieved by the corresponding \emph{fqc} configuration. Conversely, by going from a \emph{fqc} to the amorphous region, a configuration change is observed with the development of an amorphous state.These observations suggest that the amorphous phase is thermodynamically stable at large enough chirality, again similarly to the blue fog in 3D.  }

\begin{figure}[t!]
	\centerline{\includegraphics[width=0.98\columnwidth]{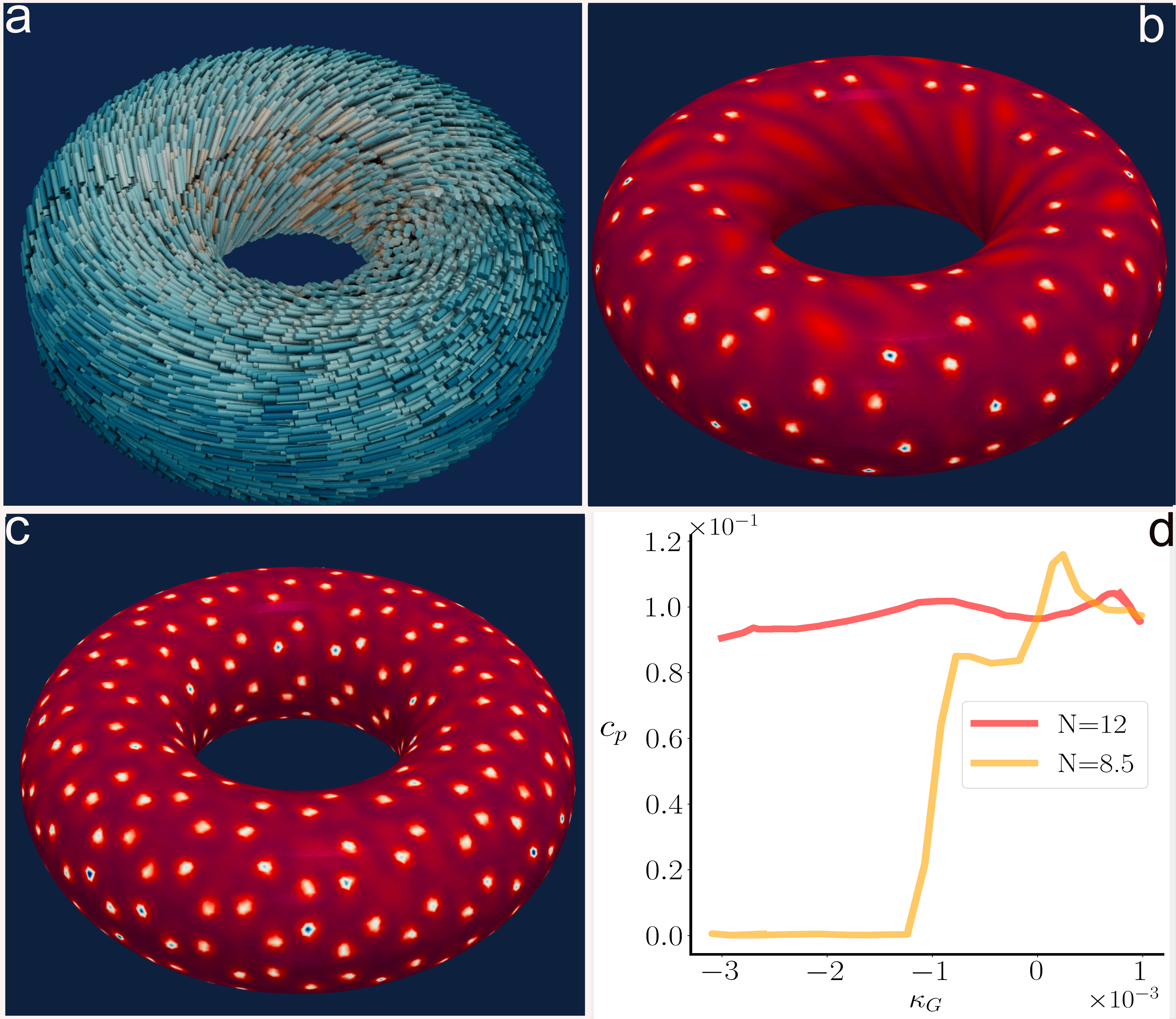}}
	\caption{\textbf{Curvature-induced topological phases. }
	(a-c) LC patterns on toroidal shells for different chirality strength $\kappa$. In the cholesteric phase of panel~(a) at $N=1$ the director field winds around the torus -- color code as in Fig.~\ref{fig1}(a). At $N=8.5$ (panel~(b)) a heterogeneous pattern emerges with a half-skyrmion lattice only in the regions with positive curvature. At $N=12$ the lattice occupies the whole surface (color code as in Fig.~\ref{fig1}(b-f)). 
	Panel~(d) shows the biaxiality parameter $c_p$~\cite{SI} against the curvature %($-0.003<\kappa_G<0.001$) 
	for the cases in panels~(b-c)}
	\label{fig5}
\end{figure}

%The full phase diagram of the system %, showing the resulting structures from our simulations found by varying $\kappa$ and $R/p$ at fixed $\tau$,
%is reported in Fig.~\ref{fig3}f. The data show that the type of \emph{fqc} observed at intermediate $\kappa$ and low $R/p$ is highly sensitive to the value of the shell radius. In general, pentagons tend to replace squares as $R$ is increased. Indeed, topologically the defect charge associated to a square is $+1/3$, while a pentagon only contributes with a $+1/6$ defect. Additionally, pentagons cover a greater area than squares and are energetically favored. Thus, in principle, each square could be substituted by two pentagons. However this is true only for large enough radii, where a large number of polygons can be arranged without introducing further frustration in the system, eventually resulting in a higher energy cost.
%\textcolor{blue}{Discutere anche blue fog}

%\section{Curvature induced N*-BP transition}
%\Livio{From here on, check after simulations are complete}
Our framework allows us to consider shells of different shapes and topology, and we discuss here the case of a toroidal surface~\cite{Ellis2018,McInerney2019} which leads to additional phenomenology (Fig.~\ref{fig5}). Since the Euler characteristic of a torus equals $0$, hexagonal half-skyrmion lattices are possible. However, unlike a flat surface a torus has variable and non-zero local curvature 
%Calling $R$ and $r$ the major and the minor radius of the torus respectively, the curvature equals $1/r(R+r)$ on the external equator, drops to $0$ at the two saddle lines and reaches its minimum negative value $-1/r(R-r)$ on the internal equator. 
%The variable curvature 
which leads to an additional space-dependent saddle-splay term in the free energy
%of a chiral liquid crystal confined to its surface. 
%The extra free energy is 
proportional to $\delta \mathcal{F}^{curv} \propto L/\kappa_G$~\cite{Sethna83} with $\kappa_G$ the local Gaussian curvature~\cite{Kamien2002}. %, of the embedded manifold~\cite{Kamien2002} (here $R_i$ represent the  curvature radii $R_i$ and $N$ is the dimensionality of the manifold).
%We find that this curvature-dependent 
Such contribution, albeit small, can strongly affect the stability and morphology of topological phases. %, which are known to be sensitive to small variations in the control parameters. %essence of the complexity of the blue-phase lies in their frustration so that a curved space may either favor their stability if $\kappa_G>0$  or prevent it otherwise.
Thus, for a fixed toroidal geometry and reduced temperature, we vary $\kappa$ by increasing $N=2R q_0/\pi$, the number of times that the director winds around the torus when in the chiral phase, being $R$ the major radius of the torus. At small $\kappa$, defect-free helical patterns are formed  %where cholesteric layers form at an angle with respect to the equatorial plane 
(Fig.~\ref{fig5}a), while topological phases are observed when chirality increases past $N\sim 8.5$.
%which are fundamentally different from those observed on the sphere. 
The curvature-dependent saddle-splay term favours half-skyrmions where $\kappa_G>0$ and the defect-free cholesteric phase where $\kappa_G<0$. Consequently, the ordering on the manifold becomes heterogeneous with a cholesteric phase in the internal region of the torus, and a hexagonal pattern of $-1/2$ defects in the outer region (Fig.~\ref{fig5}b). As expected, the boundary between the two coexisting phases forms close to the saddle lines at $\kappa_G=0$. %,  where the regular network of $-1/2$ point disclinations breaks up into an array of randomly scattered defects. 
At larger values of $\kappa$ ($N \simeq 12$) %(larger than $\simeq 1.2$, or equivalently $N$ larger than $\simeq 12$) 
the curvature-dependent perturbation is no longer sufficient to stabilise the cholesteric phase and the half-skyrmion lattice invades the whole surface (Fig.~\ref{fig5}c). 
%[The hexagonal lattice may still transiently develop dislocations consisting in pairs of pentagons and heptagons that at late times annihilate with each other through a recombination of vertices.] %, thus giving rise to a defect-free hexagonal lattice, as in a flat geometry.

%\begin{figure}[t!]
%	\centerline{\includegraphics[width=.98\columnwidth]{fig_phase_diagram.pdf}}
%	\caption{\textbf{Phase diagram}.}\label{fig3}
%\end{figure}

%\section{Conclusions}
To conclude, we investigated the nature of the topological phases arising in non-Euclidean cholesteric shells close to the isotropic-cholesteric transition. We have shown that the curved geometry, via the Gauss-Bonnet theorem, frustrates the formation of regular half-skyrmion lattices, which are instead stable on flat surfaces. %, rendering topological phases in cholesteric shells fundamentally different from those observed in 2D. 
On a spherical shell, for intermediate chirality and small radii the emerging structures are finite quasicrystals composed by a network of surface defects with topological charge $-1/2$. These structures can be seen as polyhedra composed of regular polygons, corresponding to half-skyrmion tessellations of the surface of the sphere.  %surface with global topological charge equal to the Euler characteristic of a sphere ($+2$).
For larger shells, a qualitatively distinct amorphous phase develops: this is characterized by a disordered arrangement of polygons on the shell, similar to the three-dimensional structure of blue phase III. Simulations suggest that, like blue phase III,  the amorphous phase is thermodynamically stable in a finite parameter range. The topological transition between quasicrystalline and amorphous tessellation may in some parameter range be mediated by the nucleation of dislocation scars, analogous to those found in spherical crystals~\cite{Bowick2000}.
%\Livio{This morphological transition is mediated by the proliferation of dislocation scars which locally destroy order, as the radius of the confining geometry grows}. The surface geometry also couples to the physics of the system locally, as the Gaussian curvature introduces a saddle-splay contribution in the free energy which can stabilise spatially heterogeneous structures. % with half-skyrmion lattices coexisting with helical director patterns on surfaces with non-uniform curvature, such as a torus. 

Besides being of theoretical interest, %as they provide %\st{the generalisation to} 
%a realization in terms of topological quasiparticles of the Thomson problem, %of finding the optimal arrangements of particles on a surface, 
we hope our work will also stimulate future experiments. Cholesteric shells can be created in the lab by confining liquid crystals to the surface of emulsion droplets~\cite{springer2000,higgins2005}, and surfaces with nontrivial genus can be generated. Such systems are ideally suited to search for the structures we predicted.

\section{Acknowledgments}
\begin{acknowledgments}
The work has been performed under the Project HPC-EUROPA3 (INFRAIA-2016-1-730897), with the support of the EC Research Innovation Action under the H2020 Programme. Part of this work was carried out on the Dutch national e-infrastructure with the support of SURF through the Grant 2021.028 for computational time (L.N.C and G.N.).
L.N.C. would like to thank Ireth Garcia Aguilar for useful discussions.
\end{acknowledgments}

\bibliographystyle{unsrt}
\bibliography{biblio}

\begin{thebibliography}{10}

\bibitem{bluephaselasers}
H.~Coles and S.~Morris.
\newblock Liquid-crystal lasers.
\newblock {\em Nat. Photonics}, 4, 2010.

\bibitem{skyrmionssmalyukh}
D.~Foster, C.~Kind, P.J. Ackerman, J.-S.~B. Tai, M.R. Dennis, and I.I.
  Smalyukh.
\newblock Two-dimensional skyrmion bags in liquid crystals and ferromagnets.
\newblock {\em Nat. Phys.}, 2019.

\bibitem{hopfionsmalyukh}
P.J. Ackerman, J.~van~de Lagemaat, and I.I. Smalyukh.
\newblock Self-assembly and electrostriction of arrays and chains of hopfion
  particles in chiral liquid crystals.
\newblock {\em Nat. Commun.}, 6, 2015.

\bibitem{Schwartz2018}
{M. Schwartz and G. Lenzini and Y. Geng and P.B. R{\o}nne and P.Y.A. Ryan and
  J.P.F. Lagerwall}.
\newblock {{Cholesteric Liquid Crystal Shells as Enabling Material for
  Information-Rich Design and Architecture}}.
\newblock {\em {Adv. Mater.}}, {30}({30}):{1707382}, {2018}.

\bibitem{Wright89}
D.C. Wright and N.D. Mermin.
\newblock Crystalline liquids: the blue phases.
\newblock {\em Rev. Mod. Phys.}, 61:385--432, 1989.

\bibitem{selinger}
S.M. Shamid, D.W. Allender, and J.V. Selinger.
\newblock {Predicting a Polar Analog of Chiral Blue Phases in Liquid Crystals}.
\newblock {\em Phys. Rev. Lett.}, 113:237801, 2014.

\bibitem{knotscolloidslc}
Uro{\v{s}} Tkalec, Miha Ravnik, Simon {\v{C}}opar, Slobodan {\v{Z}}umer, and
  Igor Mu{\v{s}}evi{\v{c}}.
\newblock Reconfigurable knots and links in chiral nematic colloids.
\newblock {\em Science}, 333(6038):62--65, 2011.

\bibitem{kikuchibpdevice}
Hirotsugu Kikuchi, Hiroki Higuchi, Yasuhiro Haseba, and Takashi Iwata.
\newblock 62.2: Invited paper: Fast electro-optical switching in
  polymer-stabilized liquid crystalline blue phases for display application.
\newblock In {\em SID Symposium Digest of Technical Papers}, volume~38, pages
  1737--1740. Wiley Online Library, 2007.

\bibitem{mihacolloids}
M.~Ravnik, G.A. Alexander, J.M. Yeomans, and S.~{\v Z}umer.
\newblock Three-dimensional colloidal crystals in liquid crystalline blue
  phases.
\newblock {\em Proc. Nat. Acad. Sci.}, 108(13):5188--5192, 2011.

\bibitem{henrich2011}
O.~Henrich, K.~Stratford, M.E. Cates, and D.~Marenduzzo.
\newblock {Structure of Blue Phase III of Cholesteric Liquid Crystals}.
\newblock {\em Phys. Rev. Lett.}, 106:107801, 2011.

\bibitem{kentbp3}
Sahil~Sandesh Gandhi and Liang-Chy Chien.
\newblock Unraveling the mystery of the blue fog: structure, properties, and
  applications of amorphous blue phase iii.
\newblock {\em Advanced Materials}, 29(47):1704296, 2017.

\bibitem{Fukuda2011}
{J. Fukuda and S. \v{Z}umer}.
\newblock {{Quasi-two-dimensional Skyrmion lattices in a chiral nematic liquid
  crystal}}.
\newblock {\em {Nat. Comm.}}, {2}, {2011}.

\bibitem{Nych2017}
{A. Nych and J. Fukuda and U. Ognysta and S. \v{Z}umer and I. Mu\v{s}evi\v{c}}.
\newblock {Spontaneous formation and dynamics of half-skyrmions in a chiral
  liquid-crystal film}.
\newblock {\em {Nat. Phys.}}, {13}:{1215–1220}, {2017}.

\bibitem{Metselaar2019}
L.~Metselaar, A.~Doostmohammadi, and J.M. Yeomans.
\newblock {Topological states in chiral active matter: Dynamic blue phases and
  active half-skyrmions}.
\newblock {\em J. Chem. Phys.}, 150(6):064909, 2019.

\bibitem{Fukuda2011_b}
{J. Fukuda and S. \v{Z}umer}.
\newblock {{Ring Defects in a Strongly Confined Chiral Liquid Crystal}}.
\newblock {\em {Phys. Rev. Lett.}}, {106}:{097801}, {2011}.

\bibitem{smalyukh}
J.-S.~B. Tai, P.J. Ackerman, and I.I. Smalyukh.
\newblock Topological transformations of hopf solitons in chiral ferromagnets
  and liquid crystals.
\newblock {\em Proc. Nat. Acad. Sci.}, 115(5):921--926, 2018.

\bibitem{Rokhsar86}
D.S. Rokhsar and J.P. Sethna.
\newblock {Quasicrystalline Textures of Cholesteric Liquid Crystals: Blue Phase
  III?}
\newblock {\em Phys. Rev. Lett.}, 56:1727--1730, Apr 1986.

\bibitem{Tran2017}
{L. Tran and M.O. Lavrentovich and G. Durey and A. Darmon and M.F. Haase and N.
  Li and D. Lee and K.J. Stebe and R.D. Kamien and T. Lopez-Leon}.
\newblock {{Change in Stripes for Cholesteric Shells via Anchoring in
  Moderation}}.
\newblock {\em {Phys. Rev. X}}, {7}:{041029}, {2017}.

\bibitem{Darmon2016}
{A. Darmon and M. Benzaquen and S. \v{C}opar and O. Dauchot and T. Lopez-Leon}.
\newblock {Topological defects in cholesteric liquid crystal shells}.
\newblock {\em {Soft Matter}}, {12}:{9280--9288}, {2016}.

\bibitem{Durey2020}
{G. Durey and H.R.O. Sohn and P.J. Ackerman and E. Brasselet and I.I. Smalyukh
  and T. Lopez-Leon}.
\newblock {{Topological solitons, cholesteric fingers and singular defect lines
  in Janus liquid crystal shells}}.
\newblock {\em {Soft Matter}}, {16}:{2669--2682}, {2020}.

\bibitem{Duzgun2020}
{A. Duzgun and C. Nisoli and C.J.O. Reichhardt and C. Reichhardt}.
\newblock {Commensurate states and pattern switching via liquid crystal
  skyrmions trapped in a square lattice}.
\newblock {\em {Soft Matter}}, {16}:{3338--3343}, {2020}.

\bibitem{Duzgun2021}
{A. Duzgun and C. Nisoli}.
\newblock {{Skyrmion Spin Ice in Liquid Crystals}}.
\newblock {\em {Phys. Rev. Lett.}}, {126}:{047801}, {2021}.

\bibitem{ireth2021}
I.~Garc\'{\i}a-Aguilar, P.~Fonda, E.~Sloutskin, and L.~Giomi.
\newblock Faceting and flattening of emulsion droplets: A mechanical model.
\newblock {\em Phys. Rev. Lett.}, 126:038001, Jan 2021.

\bibitem{rassegnadifettiliquidcrystalsgocce}
T.~Lopez-Leon and Fernandez-Nieves.
\newblock Drops and shells of liquid crystal.
\newblock {\em A. Colloid Polym. Sci.}, 289:345, 2011.

\bibitem{Pollard2019}
{J. Pollard and G. Posnjak and S. \v{C}opar and I. Mu\v{s}evi\v{c} and G.P.
  Alexander}.
\newblock {Point Defects, Topological Chirality, and Singularity Theory in
  Cholesteric Liquid-Crystal Droplets}.
\newblock {\em {Phys. Rev. X}}, {9}:{021004}, {2019}.

\bibitem{glotzerquasicrystals}
A.~Haji-Akbari, M.~Engel, A.S. Keys, X.~Zheng, R.G. Petschek,
  P.~Palffy-Muhoray, and S.C. Glotzer.
\newblock Disordered, quasicrystalline and crystalline phases of densely packed
  tetrahedra.
\newblock {\em Nature}, 462, 2009.

\bibitem{Carenza22065}
L.N. Carenza, G.~Gonnella, D.~Marenduzzo, and G.~Negro.
\newblock Rotation and propulsion in 3d active chiral droplets.
\newblock {\em Proc. Natl. Acad. Sci.}, 116(44):22065--22070, 2019.

\bibitem{carenza2020_physA}
L.N. Carenza, G.~Gonnella, D.~Marenduzzo, and G.~Negro.
\newblock Chaotic and periodical dynamics of active chiral droplets.
\newblock {\em Physica A}, 559:125025, 2020.

\bibitem{Metselaar2019b}
{L. Metselaar and J.M. Yeomans and A. Doostmohammadi}.
\newblock {Topology and Morphology of Self-Deforming Active Shells}.
\newblock {\em {Phys. Rev. Lett.}}, {123}:{208001}, {2019}.

\bibitem{Liu2003}
C.~Liu and J.~Shen.
\newblock A phase field model for the mixture of two incompressible fluids and
  its approximation by a fourier-spectral method.
\newblock {\em Phys. D: Nonlinear Phenom.}, 179(3):211--228, 2003.

\bibitem{Lee2012}
H.G. Lee and J.~Kim.
\newblock {Regularized Dirac delta functions for phase field models}.
\newblock {\em Int. J. Numer. Meth. Eng.}, 91(3):269--288, 2012.

\bibitem{SI}
{See online Supplemental Material at XXX for more details on the algorithm and
  additional results.}

\bibitem{callan2006}
A.C. Callan-Jones, R.A. Pelcovits, V.A. Slavin, S.~Zhang, D.H. Laidlaw, and
  G.B. Loriot.
\newblock Simulation and visualization of topological defects in nematic liquid
  crystals.
\newblock {\em Phys. Rev. E}, 74:061701, 2006.

\bibitem{alexander2009}
G.P. Alexander and J.M. Yeomans.
\newblock Numerical results for the blue phases.
\newblock {\em Liq. Cryst.}, 36(10-11):1215--1227, 2009.

\bibitem{depablo2015}
J.A. Mart{\'\i}nez-Gonz{\'a}lez, Y.~Zhou, M.~Rahimi, E.~Bukusoglu, N.L. Abbott,
  and J.J. de~Pablo.
\newblock Blue-phase liquid crystal droplets.
\newblock {\em Proceedings of the National Academy of Sciences},
  112(43):13195--13200, 2015.

\bibitem{braz1975_b}
S.A. Brazovskii and S.G. Dmitriev.
\newblock Phase transitions in cholesteric liquid crystals.
\newblock {\em J. Exp. Theor. Phys.}, 42(3):497, 1975.

\bibitem{denniston2001}
C.~Denniston, E.~Orlandini, and J.M. Yeomans.
\newblock Lattice boltzmann simulations of liquid crystal hydrodynamics.
\newblock {\em Phys. Rev. E}, 63:056702, 2001.

\bibitem{orlandini2008}
E.~Orlandini, M.E. Cates, D.~Marenduzzo, L.~Tubiana, and J.M. Yeomans.
\newblock Hydrodynamic of active liquid crystals: A hybrid lattice boltzmann
  approach.
\newblock {\em Mol. Cryst. Liq. Cryst.}, 494:293, 2008.

\bibitem{succi2001}
S.~Succi.
\newblock {\em The Lattice Boltzmann Equation: For Fluid Dynamics and Beyond}.
\newblock Numerical Mathematics and Scientific Computation. Clarendon Press,
  2001.

\bibitem{carenza2019}
L.N. Carenza, G.~Gonnella, A.~Lamura, G.~Negro, and A.~Tiribocchi.
\newblock Lattice boltzmann methods and active fluids.
\newblock {\em Eur. Phys. J. E}, 42(6):81, 2019.

\bibitem{bonelli2019}
F.~Bonelli, L.N. Carenza, G.~Gonnella, D.~Marenduzzo, E.~Orlandini, and
  A.~Tiribocchi.
\newblock Lamellar ordering, droplet formation and phase inversion in exotic
  active emulsions.
\newblock {\em Sci. Rep.}, 9:2801, 2019.

\bibitem{Metselaar2018}
L.~Metselaar, A.~Doostmohammadi, and J.M. Yeomans.
\newblock Two-dimensional, blue phase tactoids.
\newblock {\em Mol. Phys.}, 116(21-22):2856--2863, 2018.

\bibitem{Kamien2002}
R.D. Kamien.
\newblock The geometry of soft materials: a primer.
\newblock {\em Rev. Mod. Phys.}, 74:953--971, 2002.

\bibitem{SM}
See supplemental materials for further informations regarding the model,
  numerical method and further validation of the results.

\bibitem{Scacchi202}
A.~Scacchi, W.R.C. Somerville, D.M.A. Buzza, and A.J. Archer.
\newblock Quasicrystal formation in binary soft matter mixtures.
\newblock {\em Phys. Rev. Res.}, 2:032043, Aug 2020.

\bibitem{Subramanian2016}
P.~Subramanian, A.J. Archer, E.~Knobloch, and A.M. Rucklidge.
\newblock Three-dimensional icosahedral phase field quasicrystal.
\newblock {\em Phys. Rev. Lett.}, 117:075501, 2016.

\bibitem{order_parameter}
{For a regular lattice, $\psi^j_n$ does not depend on either $j$ or $n$ (in
  this case $\mathcal{N}^{j}_{1,2}=n$ and $\theta^j=2\pi/n$). For a \emph{fqc},
  $\psi_n^j$ still does not depend on $j$, but $\mathcal{N}^j_{1,2}$ may differ
  from $n$ as different types of polygons contribute to the tessellation}.

\bibitem{Note1}
For a rigorous identification we should characterise the hexatic order of these
  tessellations, which requires very large systems to achieve a good
  accuracy~\cite {digregorio2021}.

\bibitem{copar2019}
A.~Lo\v{s}dorfer Bo\v{z}i\v{c} and S.~\v{C}opar.
\newblock Spherical structure factor and classification of hyperuniformity on
  the sphere.
\newblock {\em Phys. Rev. E}, 99:032601, 2019.

\bibitem{shtools}
M.A. Wieczorek and M.~Meschede.
\newblock {SHTools: Tools for Working with Spherical Harmonics}.
\newblock {\em Geochemistry, Geophysics, Geosystems}, 19(8):2574--2592, 2018.

\bibitem{Duzgun2018}
A.~Duzgun, J.V. Selinger, and A.~Saxena.
\newblock Comparing skyrmions and merons in chiral liquid crystals and magnets.
\newblock {\em Phys. Rev. E}, 97:062706, 2018.

\bibitem{Bowick2000}
{Bowick, Mark J. and Nelson, David R. and Travesset, Alex}.
\newblock {Interacting topological defects on frozen topographies}.
\newblock {\em {Phys. Rev. B}}, {62}:{8738--8751}, {2000}.

\bibitem{Bausch2003}
{A.R. Bausch and M.J. Bowick and A. Cacciuto and A.D. Dinsmore and M.F. Hsu and
  D.R. Nelson and M.G. Nikolaides and A. Travesset and D.A. Weitz}.
\newblock {{Grain Boundary Scars and Spherical Crystallography}}.
\newblock {\em {Science}}, {299}({5613}):{1716--1718}, {2003}.

\bibitem{Irvine2010}
{W. Irvine and V. Vitelli and P. Chaikin}.
\newblock {Pleats in crystals on curved surfaces}.
\newblock {\em {Nature}}, {468}:{947–951}, {2010}.

\bibitem{Ellis2018}
{P.W. Ellis and K. Nayani and J.P. McInerney and D.Z. Rocklin and J.O. Park and
  M. Srinivasarao and E.A. Matsumoto and A. Fernandez-Nieves}.
\newblock {{Curvature-Induced Twist in Homeotropic Nematic Tori}}.
\newblock {\em {Phys. Rev. Lett.}}, {121}:{247803}, {2018}.

\bibitem{McInerney2019}
{J.P. McInerney and P.W. Ellis and D.Z. Rocklin and A. Fernandez-Nieves and
  E.A. Matsumoto}.
\newblock {Curved boundaries and chiral instabilities – two sources of twist
  in homeotropic nematic tori}.
\newblock {\em {Soft Matter}}, {15}:{1210--1214}, {2019}.

\bibitem{Sethna83}
J.P. Sethna, D.C. Wright, and N.D. Mermin.
\newblock {Relieving Cholesteric Frustration: The Blue Phase in a Curved
  Space}.
\newblock {\em Phys. Rev. Lett.}, 51:467--470, Aug 1983.

\bibitem{springer2000}
G.~H. Springer and D.A. Higgins.
\newblock {Toroidal Droplet Formation in Polymer-Dispersed Liquid Crystal
  Films}.
\newblock {\em J. Am. Chem. Soc.}, 122(28):6801--6802, 2000.

\bibitem{higgins2005}
D.A. Higgins, J.E. Hall, and A.~Xie.
\newblock {Optical Microscopy Studies of Dynamics within Individual
  Polymer-Dispersed Liquid Crystal Droplets}.
\newblock {\em Acc. Chem. Res.}, 38(2):137--145, 2005.
\newblock PMID: 15709733.

\bibitem{digregorio2021}
Pasquale Digregorio, Demian Levis, Leticia~F Cugliandolo, Giuseppe Gonnella,
  and Ignacio Pagonabarraga.
\newblock Unified analysis of topological defects in 2d systems of active and
  passive disks.
\newblock {\em arXiv preprint arXiv:2106.03454, under pubblication on Soft
  Matter}, 2021.

\end{thebibliography}
\end{document}